\documentclass[osajnl,twocolumn,showpacs,superscriptaddress,10pt]{revtex4-1} 
\usepackage{amsmath,amssymb,graphicx}
\begin{document}

\title{How to replace the oil droplet in Millikan's experiment \\with a single virus}

\author{Sanli Faez}\email{S.Faez@uu.nl}
\affiliation{Debye Institute for Nanomaterials Science, Center for Extreme Matter and Emergent Phenomena, Utrecht University, Princetonplein 5, 3584CC Utrecht, The Netherlands}

\begin{abstract}A highly sensitive optical capillary electrophoresis measurement method based on a nanofluidic optical fiber platform is presented. By using scaling arguments and considering realistic instrumental limitations, I underline the feasibility of measuring the electrophoretic mobility of a single freely-diffusing nanoparticle or macromolecule in vitro {\em continuously} and {\em indefinitely} with microsecond time resolution. The high speed of this technique opens up new possibilities for studying reaction kinetics at the single molecule level.
\end{abstract}


\maketitle 

\section{Introduction}
More than a century after Robert Millikan had reported on measuring the elementary charge, his method is still being used for experiments such as high precision detection of fractional charges at ambient conditions~\cite{mar_improved_1996} or single charging events on an optically-trapped colloid particle in an apolar solvent~\cite{strubbe_detection_2008, beunis_beyond_2012}. The main reason for the longstanding avail of this method is probably its elegant simplicity: by observing the drift velocity of a particle forced by an external electric field one obtains its electrical mobility, which has usually a well-defined monotonic relation with the net charge of the particle. While metrological descendants of Millikan's apparatus are mainly used for detecting stable fractional charges~\cite{lee_automated_2004}, closely related experiments have been realized during the last decade for measuring the net charge of a mobile \emph{single} nanoparticle~\cite{mojarad_measuring_2012} or charge dynamics of a single molecule \cite{goldsmith_redox_2011, wang_single-molecule_2014} in aqueous liquids. It needs no imagination to grasp the significance of reaching this goal. If one can monitor the charge (or the electrophoretic mobility) of a single solute rapidly enough, she or he will be able to study kinetic interactions such as ionization, hydrolysis, or charge transfer at the single particle level. In such an experiment, one can directly visualize the intermediate steps of a reaction, which might be untraceable in bulk experiments due to the sheer magnitude of the Avogadro's number.
\begin{figure}[htbp]
\centerline{\includegraphics[width=.8\columnwidth]{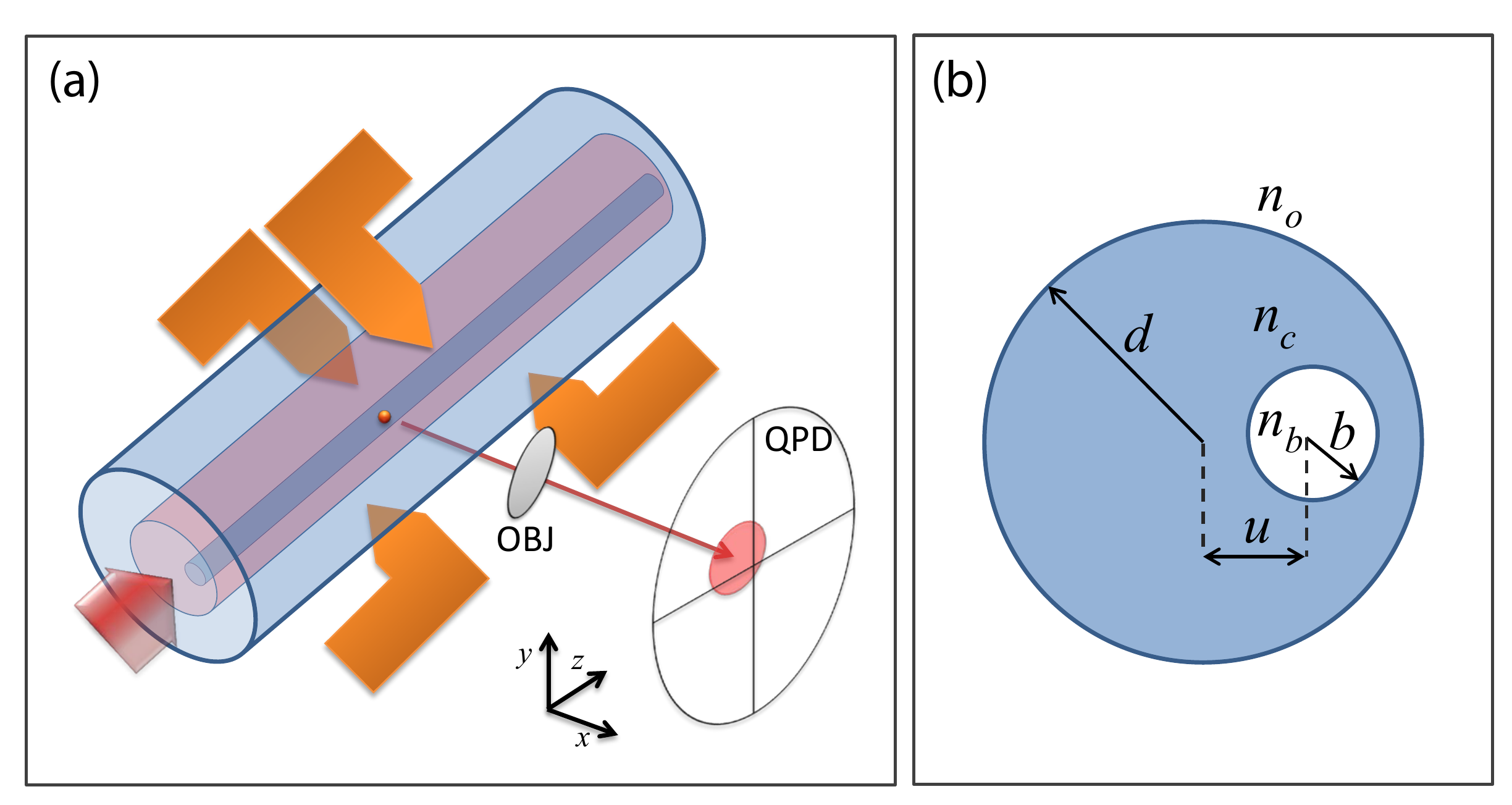}}
\caption{\label{fig1setup}Schematic of the proposed setup for detection and feedback trapping a single nanometric object with elastic scattering of guided light. The nanoparticle is diffusing inside a nanofluidic optical fiber and the out of axis scattering signal from the particle is collected with a microscope objective (OBJ) and directed to a quadrant photo diode (QPD). The difference signals of the QPD indicate the particle displacements in $y$ and $z$ directions while the total intensity is representative of the $x$-coordinate. The four orange arrows represent the metallic electrodes that are used for applying an arbitrary external electric field to counter thermal drift. (b) The cross-section of the asymmetric hollow core of the silica fiber that is required for this trapping platform.}
\end{figure}

Monitoring the mobility of single nanoparticles (smaller than 100 nm) in dense fluids continuously is difficult for two reasons: 1-Brownian motion and 2-diminishing optical visibility. It is straightforward to prove that the diffusion due to Brownian motion is not a fundamental challenge, as long as one does not lose the particle under investigation. Although smaller particles diffuse faster, they also exhibit a higher mobility. In fact, while the displacement due to an external uniform driving force changes linearly with time, the uncertainty caused by Brownian motion increases only proportional to the square root of time. Therefore, it is easy to overcome this uncertainty if one actuates the particle strongly enough. Optically tracking the particle motion, however, is the single most challenging element for mobility measurement at the single particle level since the optical signal (for example scattering intensity) from sub-wavelength particles decreases steeply as the particle size gets smaller. This decrease is proportional to the particle’s volume when using the fluorescence signal (for a given dye concentration) or scales with volume squared when the elastic scattering is used for detection. Despite the steeper scaling elastic scattering is more compatible with high-flux illumination because it is non-dissipative. The fluorescence emission is on the other hand prone to saturation and photo-bleaching under strong illumination. This limitation on optical signal is even more pressing in high-speed measurements because of the photon shot noise; i. e. one needs a minimum amount of photons per frame to overcome the dark-noise.

In a breakthrough experiment, Cohen and Moerner~\cite{cohen_suppressing_2006} have developed the anti-Brownian electrokinetic (ABEL) trap based on fluorescent microscopy and dynamic suppression of the particle displacement by electrophoretic forces. The feedback signal in the ABEL trap is also used to determine the dynamic electrophoretic mobility in real time. More recently, Wang and Moerner~\cite{wang_single-molecule_2014} have further improved this method and extracted both the mobility and the diffusivity of a single protein with a time resolution of 200~ms using time-stamped single photon counting. Despite these remarkable achievements, the time resolution of such measurements in the ABEL trap seems hard to improve further because of the limited fluorescence signal~\cite{kumbakhar_feedback_2014} and the duration of the measurement is truncated by photobleaching of the fluorescent labels.  These limitations can in some cases be overcome by using elastic light scattering, instead of fluorescence, for tracking the solute particle. Mojarad and Krishnan~\cite{mojarad_measuring_2012} have used video microscopy based on inline interferometric scattering (iSCAT) detection to track an 80-nm gold nanoparticle inside an electrostatic trap~\cite{krishnan_geometry-induced_2010} and determine its bare charge. This particular measurement has been done passively by probing the probability distribution of the particle position over hundreds of frames and comparing it with the simulated static trap potential, considering that the higher charged particle are trapped more tightly. I suggest that by combining this detection method with feedback-controlled trapping, the time-resolution of this method in measuring dynamic electrophoresis can still be improved. However, position detection with iSCAT is based on recording an image and making a fit to the expected diffraction pattern, which will ultimately limit the feedback latency~\cite{cohen_thesis_2006}.

In this article, I propose a fiber-based method based on elastic scattering detection that allows high-speed three-dimensional tracking of small dielectric objects with a single quadrant photodiode (QPD). The practical time-resolution of this method is estimated to be 12~$\mu$s, {\em under continuous wave illumination}, for a 10-nm gold nanoparticle and to to be enhanced proportional to the (diameter)$^\frac{5}{2}$ for larger particle, as far as the electronic bandwidth of the detector allows. This method is based on our recent experimental achievement in tracking small virions and dielectric nanoparticles inside a nano-fluidic optical fibers with video microscopy~\cite{faez_tracking_2015}. Here, a different fiber design is presented that allows unambiguous coordinate determination with a fast QPD. This improvement is crucial for realizing the sufficient feedback bandwidth that is necessary for stable electrokinetic trapping inside the fiber. I present one optimized design for the opto-fluidic fiber and discuss the technical requirements for high-speed monitoring of the single particle mobility based on this platform.

\section{Scaling considerations}
Consider the model system of a spherical particle with radius $a$ and bare charge $q$ freely diffusing in a deionized solvent with dynamic viscosity $\eta$. The influence of charge screening and electrokinetic effects will be discussed later, but it does not influence the generality of the following argument. Under an externally applied electric field $E$, The average displacement increases linearly with time; ${\Delta x}=\mu E \Delta t$ where $\mu$ is the electrical mobility and $\Delta t$ is the time between two successive position measurement. Brownian motion, causes a deviation from displacement at constant velocity given by $\delta x = \sqrt{2 D \Delta t}$ with $D$ the translational diffusivity constant. Diffusivity and mobility are related to each other by the Einstein's relation $D=\mu kT/q$. For given values of $E$ and $q$, the displacement goes linearly with the product $\mu \Delta t$ while the dynamic error due to thermal motion grows proportional to its square root. Therefore, when considering thermal diffusion as the sole dispersion factor in the displacement measurement, the mobility (charge) of smaller particles can, in principle, be measured faster than larger ones with the Millikan's method.

Another source of uncertainty, which is more prominent in high-speed measurements, is the optical localization error mainly limited by the photon shot-noise. Generally, the lowest optical uncertainty in localizing the position of a point source~\cite{betzig_proposed_1995} (a very good approximation for a sub-wavelength nanoparticle) is given by $\delta s \approx \lambda/2 \sqrt{N}$ where $\lambda$ is the optical wavelength and
\begin{equation}\label{eq_nphotons}
N=\frac{f_R I\sigma_s}{h\nu} \tau
\end{equation}
is the number of detected photons during a given exposure time $\tau$. Here, $I$ is the illumination intensity, $\sigma_s$ is the scattering cross section, $h\nu$ is the energy of a single photon, and $f_R$ is the ratio between the number of detected to scattered photons, which represents the instrument response function and the detector quantum efficiency. As we see, the localization error $\delta s$ decreases and the dynamic error $\delta x$ increases with the square root of exposure time. Under continuous wave (CW) illumination at a fixed intensity, the optimum exposure time is at the point where these two uncertainties are equal, which reads
\begin{equation}\label{eq_optimum}
\tau_o=\frac{\lambda\sqrt{h\nu}}{\sqrt{8 R D I\sigma_s}}.
\end{equation}
For spherical objects in a viscous fluid $D \propto a^{-1}$ and for particles smaller than a wavelength $\sigma_s \propto a^{6}$ considering the Rayleigh scattering. Therefore, the optimum exposure time scales as $\tau_o \propto a^{-\frac{5}{2}}$. Not that that by using stroboscopic (pulsed) illumination, the dynamic error can be completely suppressed and the time resolution is then merely limited by the highest viable illumination intensity.

\begin{figure}[htbp]
\centerline{\includegraphics[width=.8\columnwidth]{{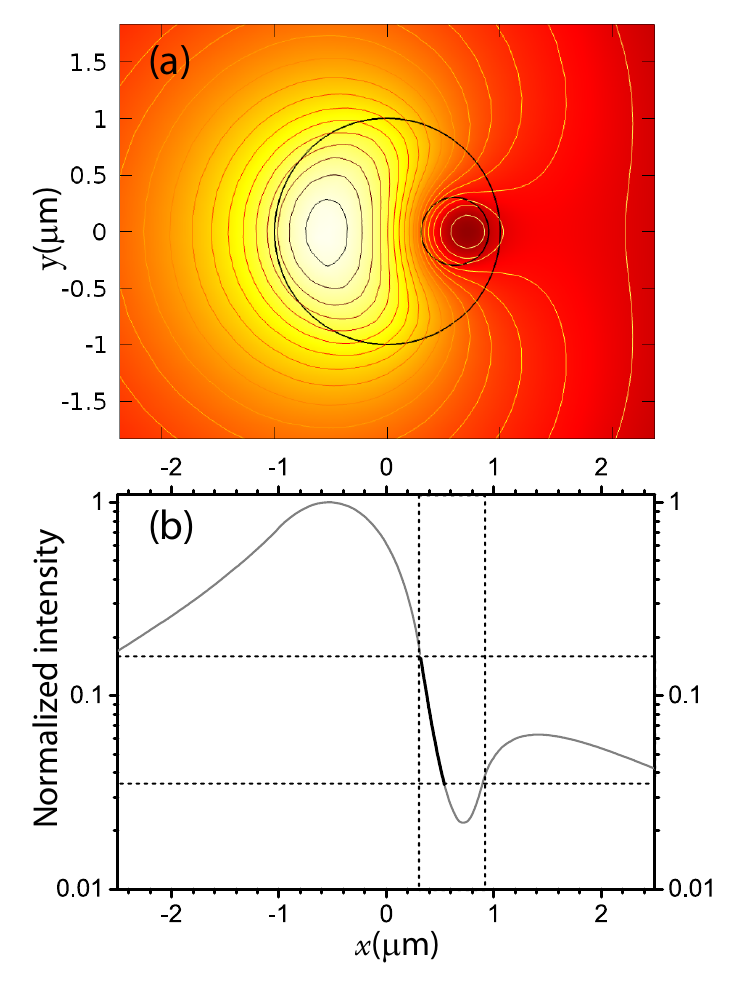}}}
\caption{\label{fig2mode}(a)False-color image and contour plot of the electric field magnitude in the y-polarized mode of the nanofluidic fiber at $\lambda=670$~nm. Parameters: $n_o=1.447$, $n_c=1.452$, $n_b=1.33$, $d=1000$~nm, $b=600$~nm, $u=600$~nm. (b) Semilogarithmic plot of the square electric field (normalized to its maximum) on the $y=0$ line. The vertical dashed lines indicate the boundaries of the inner nanochannel.}
\end{figure}

\begin{table}[htbp]
  \caption{\label{tblsize}Characteristics of spherical objects with $\sigma_s=10^{-2}\;\mathrm{nm}^2$ in water at 670~nm vacuum wavelength.}
  \begin{center}
    \begin{tabular}{lll}
    \hline
    Material & Diameter (nm) & Diffusivity\footnote{Bulk diffusion in water at 293 K, $\eta=1.0$~cP} ($\mu \mathrm{m^2/s}$)\\
    \hline
    Gold & 10.6 & 40.4 \\
    TiO$_2$ & 15.6 & 27.5 \\
    Polystyrene & 24.3 & 17.6 \\
    Protein\footnote{considering an average refractive index of 1.45} & 31.2 & 13.7 \\
    \hline
    \end{tabular}
  \end{center}
\end{table}

\section{Methods}
The proposed experimental setup is sketched in Fig.~\ref{fig1setup}(a). The central part of this setup is the nanofluidic optical fiber with the cross-section depicted in Fig.~\ref{fig1setup}(b). Light is guided in the high index core of the optical fiber that supports a single optical mode for each polarization. A sub-wavelength open channel is embedded inside the core off-center to the axis and is extended along the length of the fiber. This channel should be filled with the liquid that contains the solute particles. Out of axis scattering of light by the particle is collected by a high-NA microscope objective and focused on a QPD. The exterior of the fiber is embedded in index-matching oil to minimized the cylindrical abberations caused by the cladding. The principles of detection are similar to Ref.~\cite{faez_tracking_2015}, except for the difference in fiber geometry and use of a QPD instead of a camera. The particle displacement on the two directions orthogonal to the detection optical axis, $y$- and $z$-coordinates, can be measured by the difference signals from the QPD. A crucial difference of this design with the fiber used in Ref.~\cite{faez_tracking_2015} is in the asymmetric positioning of the central channel with respect to the core, which generates an asymmetric optical mode. This asymmetry allows for locating the particle along the detection axis based on the sum signal of the QPD.

I have used the COMSOL multiphysics package to simulate the optical modes. A typical example of the mode profile is presented in Fig.~\ref{fig2mode} considering the channel is filled with water. The mode polarization is chosen vertical in the depicted coordinates so that the in-plane scattered light can be maximally collected by the objective. Because of the lower refractive index of water relative to the silica core, the mode intensity decays exponentially close to the channel walls and reaches a minimum somewhere in the center of the channel. The value of the normalized mode intensity on the $y,z=0$ axis is depicted in Fig.~\ref{fig2mode}(b). A part of this curve is highlighted by a thick line to emphasize the interval where the scattered intensity uniquely determines the $x$-coordinate of the particle on this axis. The vertical and horizontal difference signals and the sum signal from the QPD are sufficient to determine the position of the particle in the region close to the inner wall of the channel. These signals are used as feedback to generate the necessary electric field by four electrodes outside the fiber for opposing the instantaneous drift of the particle.

In the following, I choose some typical parameters to show the effectiveness of the proposed method. I consider a particle with scattering cross section of $\sigma_s = 10^{-2}\;\mathrm{nm}^2$. This value is roughly half the absorption cross section of a single rhodamine 6G molecule~\cite{soper_photophysical_1993}. The corresponding size and diffusivity of such particles, composed of various materials, are listed in Table~\ref{tblsize}. I note in passing that for smaller particles, one can obtain a similar cross section by adding fluorescent dyes to the particle in order to increase its polarizability. Here, the detection can still be based on total scattering (elastic and inelastic) without a need for optical filter. By choosing the excitation frequency on the red-side of the dye emission spectrum, one can also supress the chance of photobleaching.

Consider a particle of $2a=10$~nm made of gold. The diffusivity of this particle in water at room temperature is roughly $40\;\mu \mathrm{m^2/s}$. To obtain the optimum exposure time from Eq.~\ref{eq_optimum} I take $\lambda=670$~nm, an overall collection efficiency of $R=0.01$, and a relatively high but practical light intensity of 30~$\mathrm{mW/\mu m}^2$ at the position of the nanoparticle inside the channel. The time-bin (for CW illumination) then reads $\tau=12~\mu\mathrm{s}$ and the number of detected photons at each interval is $N=120$ per time bin. This value means that the particle can be tracked at a remarkably high rate of more than 80~kHz. To achieve a stable feedback trap with the fiber mode depicted in Fig.~\ref{fig2mode}(b) the particle should never diffuse out of the high intensity part of the mode, i.e. $\Delta x < 100$~nm. To fulfil this condition for the chosen particle size a minimum feedback bandwidth of 8~kHz is required, which can be easily achieved with the currently available technology and the considerably high signal to noise ratio in this method.

\section{Discussion}
Unlike electrostatic traps~\cite{krishnan_geometry-induced_2010}, the performance of the trapping method presented here does not depend on the ionic strength of the solvent. However, it is still necessary to prevent the solute from sticking to the inner walls of the channel. For negatively charge particles in an aqueous solution, this is an straightforward requirement since the silica surface also obtains negative charge in contact with water. For some other solutes, such as proteins, it might be necessary to coat the channel walls with hydrophobic layers as has been done for ABEL trap~\cite{goldsmith_redox_2011}.

The elastic scattering detection used here is also instantaneous, unlike fluorescence detection that is based on spontaneous emission. Therefore, it is possible to use pulsed techniques such as pump and probe measurements to study, deterministically, the particle motion at much shorter time scales down to a few femtoseconds. This will open up the possibility of probing the ballistic motion of small particles that could not be investigated by methods based on optical traps~\cite{li_brownian_2013}. Furthermore, additional types of spectroscopy, such as Raman or broadband scattering, can be performed on the trapped particle through using the same optical illumination and detection paths.

\section*{Acknowledgements}
Fruitful discussions with John Crocker, Howard Stone, and Vinothan Manoharan are acknowledged. This article is dedicated to the memory of Harvey Fletcher.

\bibliographystyle{ieeetr}
\bibliography{nanoMillikan}


\end{document}